\documentclass[runningheads]{llncs}

\usepackage{eccv}

\usepackage{eccvabbrv}

\usepackage{graphicx}
\usepackage{booktabs}
\usepackage{subcaption}
\usepackage{multirow}
\usepackage{mathrsfs}
\usepackage{float}
\usepackage{placeins}
\usepackage{amsmath} 
\usepackage{xcolor}

\usepackage[accsupp]{axessibility}  

\usepackage{hyperref}

\usepackage{orcidlink}

\begin{document}

\title{Closing the Real-to-Synthetic Gap with SNIC: High-Fidelity Noise Modeling and Calibration} 

\titlerunning{Closing the Real-to-Synthetic Gap with SNIC}

\author{Nik Bhatt\inst{1} \and
Gail Kaiser\inst{2}\orcidlink{0000-0002-8791-1178}}

\authorrunning{N.~Bhatt et al.}

\institute{Columbia University, New York, NY\\
\email{n.bhatt@columbia.edu}\\
\and
Columbia University, New York, NY\\
\email{kaiser@cs.columbia.edu}}

\maketitle

\begin{abstract}
  Training advanced denoising models requires large datasets of high-fidelity, physically accurate images. While heteroscedastic noise models can simulate realistic noise, methodologies for their calibration remain under-explored, and large-scale calibrated datasets are scarce. We present a rigorous calibration and tuning pipeline for building high-quality heteroscedastic noise models across a range of sensors, incorporating dark frames to capture signal-independent noise. When evaluated with a state-of-the-art denoiser, our synthesized noisy RAW images reduce the Peak Signal to Noise Ratio (PSNR) gap to real-world noise by 54-64\% compared to synthesized RAW images created using manufacturer-provided noise profiles, which fail to account for smartphone ISP processing that suppresses noise in RAW files during calibration. Leveraging our pipeline, we introduce the Synthesized Noisy Images using Calibration (SNIC) dataset: over 6600 images across 30 scenes and four sensors (DSLR, point-and-shoot, and smartphone), with open-source calibration code and noise models. To our knowledge, SNIC is the only publicly available dataset with calibrated synthesized noise providing paired RAW and TIFF data, offering a new resource for researchers developing noise reduction models.
  \keywords{Denoising \and Noise Modeling \and Image Dataset}
\end{abstract}

\section{Introduction}
\label{sec:intro}

All digital images contain noise as a by-product of the capture process.
Noise arises from the quantum effects of photon arrival, and from
various electronic components in the sensor and camera circuitry,
including sensor readout, amplifiers, and analog-to-digital conversion.
Images captured in low-light environments, and especially with the
smaller sensors in smartphones, typically include considerable
amounts of noise. This degrades image quality. Thus, noise reduction, or denoising, is an important and
difficult task that has been the focus of active research for years. Denoising is best
performed on RAW images (rather than camera-generated JPEGs) because RAW
images are scene-referred, contain linear data, and undergo less processing by the image signal
processor (ISP).

Currently, most research in denoising focuses on machine learning models.
Thus, researchers developing new denoising algorithms rely on large
image datasets to train and test their models. There are three main
approaches to acquire an adequate dataset.

\begin{enumerate}
\item
  Add noise to a generic dataset (such as Flickr2K) using Additive White Gaussian Noise (AWGN). AWGN is a simple homoscedastic model which lacks physical realism and cannot accurately simulate noise from digital sensors.
\item
  Use a dataset with real noisy images \cite{SID,SIDD,RENOIR,DND,ELD,MIDD,PolyU}. Most were created in 2018 or earlier, using older cameras such as the iPhone 7. These datasets are often limited to specially lit indoor scenes.
\item
  Use a heteroscedastic Poisson-Gaussian noise model which can be easily retrieved from Digital Negative Specification (DNG) \cite{AdobeDNG} images. However, we found those models are often inaccurate, particularly for iPhones, where we discovered ISP content-aware smoothing greatly affects calibration data, requiring a novel tuning approach. As a result, we have chosen to create custom physically-based heteroscedastic noise models by calibrating cameras and refining the models with special tuning images. This approach requires more effort, but our models provide perfectly aligned, high-quality noisy RAW images with full control over the choice of cameras and scenes.
\end{enumerate}

The choice and quality of the training dataset directly affects the performance
of a denoising model on real images. For physically modeled
datasets, the key is minimizing the real-to-synthetic gap --- the discrepancy between the synthetic
noise and actual noise \cite{Zhang}. In this paper, we focus on physics-based
noise modeling, using careful calibration to determine photon shot
noise, combined with complete capture of signal-independent noise
sources via dark frames. We confirm the inaccuracy and incompleteness of manufacturer DNG
noise profiles. We demonstrate how to create and tune a heteroscedastic,
physically-based noise model to produce accurate synthesized noise, and
use our models to create a large synthesized noisy image dataset, with
both RAW and TIFF images. Our noisy images produce excellent results
both in LPIPS measurements and in denoising tests, closing the real-to-synthetic gap.

Our main contributions are as follows:
\begin{itemize}
\item
  Steps for accurately calibrating different types of cameras, from
  smartphones to DSLRs, and tuning the resulting heteroscedastic models.
\item
  Identification of iPhone content-aware smoothing in flat-field RAW images.
\item
  Heteroscedastic noise models that achieve a 54\% to 64\% reduction
  in PSNR gap vs.~standard DNG noise models.
\item
  An adaptive synthesis strategy combining calibrated P-G models with dark current frames for sensors where Gaussian read noise is insufficient to represent signal-independent noise.
\item
  A novel dataset (SNIC) containing 6602 paired RAW and TIFF images across 30 scenes. The dataset is located on the Harvard Dataverse at \url{https://doi.org/10.7910/DVN/SGHDCP}. SNIC is the only published dataset providing perfectly aligned synthesized noisy RAW images via a rigorous calibration pipeline specifically accounting for mobile ISP smoothing.
\item
  Source code to calibrate, tune, evaluate, and generate high-quality
  noisy RAW and TIFF images. The code can be found on GitHub at \url{https://github.com/nikbhatt-cu/SNIC}.
\end{itemize}

\section{Related Work}\label{related-work}

This section discusses the principal ways in which noise
models and datasets are created to train machine learning denoising
models. While learning-based noise generation methods exist \cite{Noise_Flow},
this paper focuses on physics-based statistical approaches for dataset
creation, and we do not discuss denoising models in depth.

Additive White Gaussian Noise (AWGN) adds
noise to all pixels using a simple Gaussian
distribution with zero mean and a chosen variance \cite{Noise_Flow}. 
While easy to implement, AWGN unrealistically treats noise as purely
signal-independent. Poisson-Gaussian (P-G) is a physics-based noise model which improves
upon AWGN by including both a signal-dependent component (shot noise),
and a signal-independent component (read noise). Shot noise occurs
because photon arrival is a quantum random process. For a pixel with
mean signal \(\mu\), the actual observed photon count includes random
photon shot noise \(N_p\). The total photon count follows a Poisson
distribution \cite{Zhang}: 
\begin{equation}
(N_p + \mu) \sim \mathcal{P}(\mu),
\end{equation}
where
\(\mathcal{P}\) is the Poisson distribution with mean and variance
\(\mu\). Signal-independent read noise is generated by the sensor's
readout, ADC converters and other circuitry. This noise appears even in
the absence of light. Read noise is modeled using a Gaussian
distribution with zero mean.

A P-G model includes two terms, the signal-dependent Poisson scaling
term ($a$) and the signal-independent Gaussian bias term ($b$). Because creating a
complete P-G noise model is painstaking work, researchers typically use
preexisting noise profiles extracted from DNG RAW images.

\subsubsection{DNG Noise Profiles}\label{dng-noise-profiles}

Adobe, Inc.~is the creator of DNG, which is an open format for encoding digital images \cite{AdobeDNG}. The DNG NoiseProfile tag (51041)
uses the Poisson-Gaussian formulation. There are some weaknesses in the
NoiseProfile specification. First, the specification simply contains one
or more pairs of numbers (the scaling and bias parameters from
Poisson-Gaussian). It does not include the ISO for which the noise
applies, and there is no way to provide the parameters for multiple ISO
values. Second, while some DNG NoiseProfiles include the parameters for
each of the Bayer pattern channels (R, G1, G2, and B), it is very common
to find NoiseProfiles that contain only three pairs of numbers (R, G,
B), or simply a single pair of numbers which are then applied to all
four channels equally (iPhone DNGs contain only this single pair). Since
the ISO is not specified in the NoiseProfile tag, one must examine DNG
files at different ISOs to determine the full DNG noise model.

Finally, beyond the lack of per-channel information, DNG noise profiles
provided by camera manufacturers tend to be inaccurate compared
to calibrated profiles \cite{Zhang}. We have found that profiles generated by
Adobe for proprietary RAW files (such as the Sony ARW format) can be
more accurate, but still retain the other issues described above.

\subsubsection{Calibration and Dark
Current}\label{calibration-and-dark-current}

Modeling noise accurately using a Poisson-Gaussian formulation requires
careful calibration (described in Methods). Healey and Kondepudy~\cite{Healey} discuss
calibrating a CCD using a gray card, as well as extracting Fixed Pattern
Noise by shifting both the light source and gray card. In our
calibrations, we confirmed the finding by Zhang \etal~\cite{Zhang} that fixed pattern
noise is not a relevant component in modern CMOS sensors. Thus, our
method for calibration deviates somewhat from Healey and Kondepudy\cite{Healey} and includes a
tuning portion using real noisy images at a range of ISO
levels.

The signal-independent part of noise appears in all images from a given
sensor, even completely dark images (\eg with a lens cap on). Any
electronics component can produce noise \cite{Zhang}; as a result, it is
extremely difficult, if not impossible, to fully characterize and model
this noise physically. Zhang \etal~\cite{Zhang} provided the complete formulation for a
digital image $D$: 
\begin{equation}
D =(K_a(I+ N_p + N_1) + N_2 + N_q )K_d,
\end{equation} where $I$ is the incident light signal, $N_p$ is photon shot noise as defined above,
\(K_a\) and \(K_d\) represent the analog and digital gains, \(N_1\) includes
noise affected by analog gain, \(N_2\) includes thermal and column fixed
pattern noise that takes place after the analog gain step, and \(N_q\)
is quantization noise when converting analog values to digital ones.

There are a few techniques for capturing signal-independent noise for
the purpose of noise synthesis. The simplest (and most common) is to
compute it as part of the linear regression required for P-G noise
modeling. The second is to capture dark frames (frames with no light
hitting the sensor, such as with a lens cap, and/or in a completely dark
room). This second approach, described by Zhang \etal~\cite{Zhang} as dark current, is
used in lieu of the Poisson-Gaussian bias term. It more accurately
captures sensor-independent noise, as dark images include all noise
sources by definition. However, as different sensors (and camera
electronics) produce different amounts of noise (and at different
stages), we have found that the value of capturing dark current varies
substantially from sensor to sensor. Specifically, professional-quality
cameras can be very accurately modeled using a standard P-G linear
regression. Further, some image signal processors (ISPs) in cameras,
such as the iPhone 11 Pro, will process RAW images to suppress or alter
noise based on the contents of the scene.

\subsubsection{Denoising Datasets}\label{denoising-datasets}
Published datasets fall into two categories: ones using AWGN, and ones using real noisy images. P-G datasets are not published because researchers can generate them from DNG noise profiles --- though as we show, DNG profiles are often inaccurate and incomplete. Due to the quality issues with AWGN, and the lack of published P-G datasets, we only list real noise datasets. Table~\ref{tab:datasets} lists some
popular real noisy image datasets.

\setlength{\tabcolsep}{8pt}
\begin{table}[t]
  \scriptsize
  \caption{Real Noise Datasets. Note: Most RAW datasets listed are from 2017-2018, highlighting a need for updated datasets.}
  \centering
  \begin{tabular}{l l r r l l r}
  \toprule\noalign{}
    \textbf{Dataset} & \textbf{Year} & \textbf{Images} & \textbf{Scenes} & \textbf{Cameras} & \textbf{Format} & \textbf{Citation} \\
   \midrule\noalign{}
    DND & 2017 & 50 & 50 & 4 DSLRs  & RAW & \cite{DND}\\
    SID & 2018 & 5094 & 424 & 2 DSLRs & RAW & \cite{SID}\\
    SIDD & 2018 & 30000 & 10 & 5 Phones & RAW & \cite{SIDD} \\
    RENOIR & 2018 & 480 & 120 & 3 Mixed & RAW & \cite{RENOIR} \\
    ELD & 2020 & 640 & 10 & 4 DSLR  & RAW & \cite{ELD}\\
    PolyU & 2018 & 80 & 40 & 5 DSLR & JPEG & \cite{PolyU}\\
    MIDD & 2024 & 420000 & 50 & 20 Phones & PNG & \cite{MIDD}\\
    \bottomrule\noalign{}
  \end{tabular}
  \label{tab:datasets}
\end{table}

Real noise datasets group low-ISO clean images with noisy images
captured at various ISO values \cite{ELD}. Perfect alignment is
challenging; even sub-millimeter camera movement can cause misalignment.
Further, strict lighting requirements typically limit them to indoor
scenes. For example, SIDD contains 30000 images but only 10 distinct
scenes \cite{ELD}, while DND provides just 50 image pairs \cite{DND}. Most
real noise datasets were created in 2017-2018, predating recent sensor
advances. Perfectly aligned real noise is a high bar, but
it provides an unimpeachable ground truth. Some datasets,
including MIDD \cite{MIDD} and PolyU \cite{PolyU}, only include JPEG or PNG images, so
they are not usable for RAW-based denoising.

A good alternative to real noise is Poisson-Gaussian synthesis. DNG RAW images include P-G noise profiles
provided by the manufacturer, or by Adobe as part of its camera
calibration process. Because the quality of P-G models varies greatly,
we created our own high-quality noise models. While somewhat laborious
to create, our models have advantages over real noise. Our noisy images
are perfectly aligned by definition, both indoor and outdoor scenes can
be captured with a variety of lighting, and there is no limit to the
number of images that can be produced. Further, by using dark current
frames when appropriate, complex noise sources can be modeled.

\section{Methods}\label{methods}

\subsection{RAW Images and Noise}\label{raw-images-and-noise}

There are two principal ways to process
images for noise modeling and synthesis: render the Bayer data into RGB data, or work with the Bayer data directly. We do the latter because Bayer data is linear, preserves the two green channels, and avoids dependence on specific demosaic algorithms or tone curves. Further, a RAW-based noise model can produce different values for each of the two green channels, which is not possible once they have been processed into RGB.

\subsection{DNG Noise Profile
Analysis}\label{dng-noise-profile-analysis}

The Poisson-Gaussian formulation enables straightforward synthesis of
noisy images. One can build a noise model manually, or rely on published ones. Because building models can be labor-intensive and
error-prone, most researchers rely on ones stored in DNG files. For proprietary RAW files, one can convert RAW files into DNGs using Adobe's DNG Converter software and read the NoiseProfile tag. This noise model is computed by Adobe
as part of its process for supporting proprietary RAWs. The DNG NoiseProfile specification does not
include ISO data, so to build an accurate P-G model, one must provide
the DNG Converter a series of RAW files shot with the desired range of
ISO values. We found that for two Sony cameras (RX100 IV and A7R III),
the Adobe profiles are very similar to the ones we built through our
calibration and include noise parameters for red, green, and blue
channels (but not for both green channels).

\begin{figure}[t]
  \centering
    \includegraphics{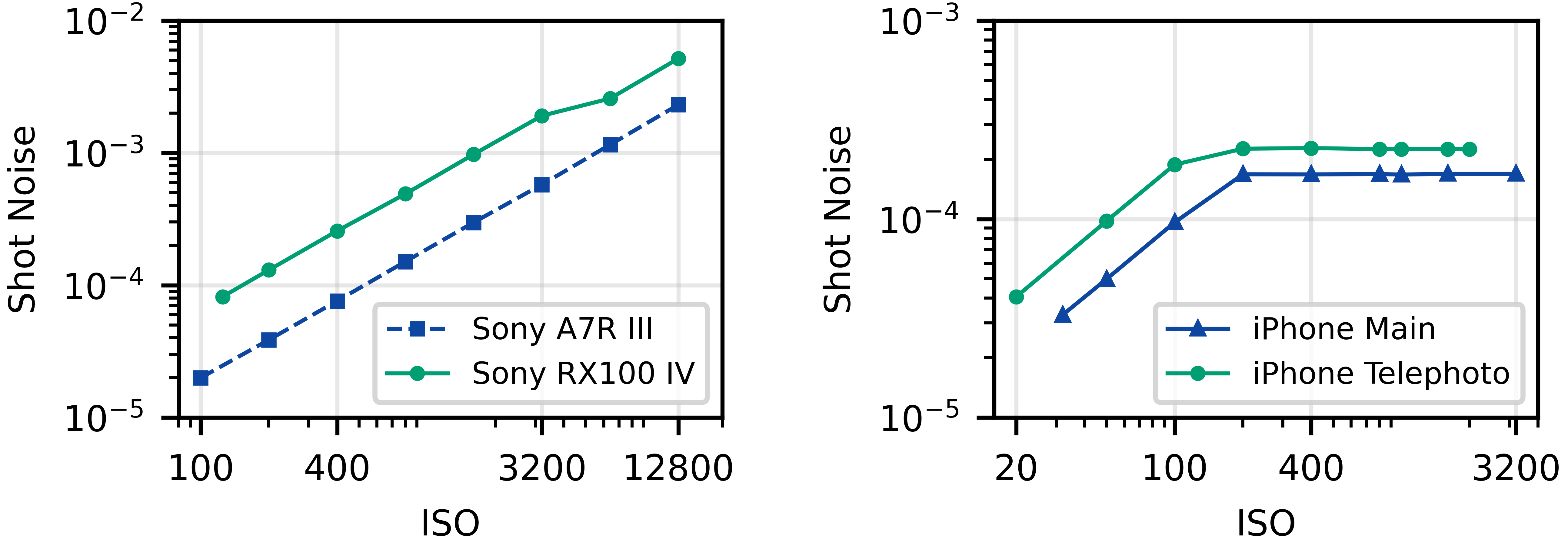}
  \caption{Comparison of DNG Profiles. Note: Y-axis scales differ due to varying sensor characteristics. iPhone noise parameters exhibit non-linearity at high ISO values.}
  \label{fig:sony_iphone}
\end{figure}

For cameras that produce DNG files
natively (such as smartphones), noise profiles are provided by the manufacturer, not Adobe. We found the DNG noise profile for the iPhone 11 Pro is both
incomplete and fundamentally flawed, especially at higher ISO values, which
confirms the findings of Zhang \etal~\cite{Zhang}. First, the DNG NoiseProfile for the
iPhone 11 Pro does not include scaling and offset values for each
channel, providing only a single pair of values. Second, as the
ISO rises, the scaling factor should rise, since noise is related to
analog gain (ISO). However, as shown in \cref{fig:sony_iphone} for the iPhone 11 Pro, the values level off.

\begin{figure}[t]
  \centering
    \includegraphics
    {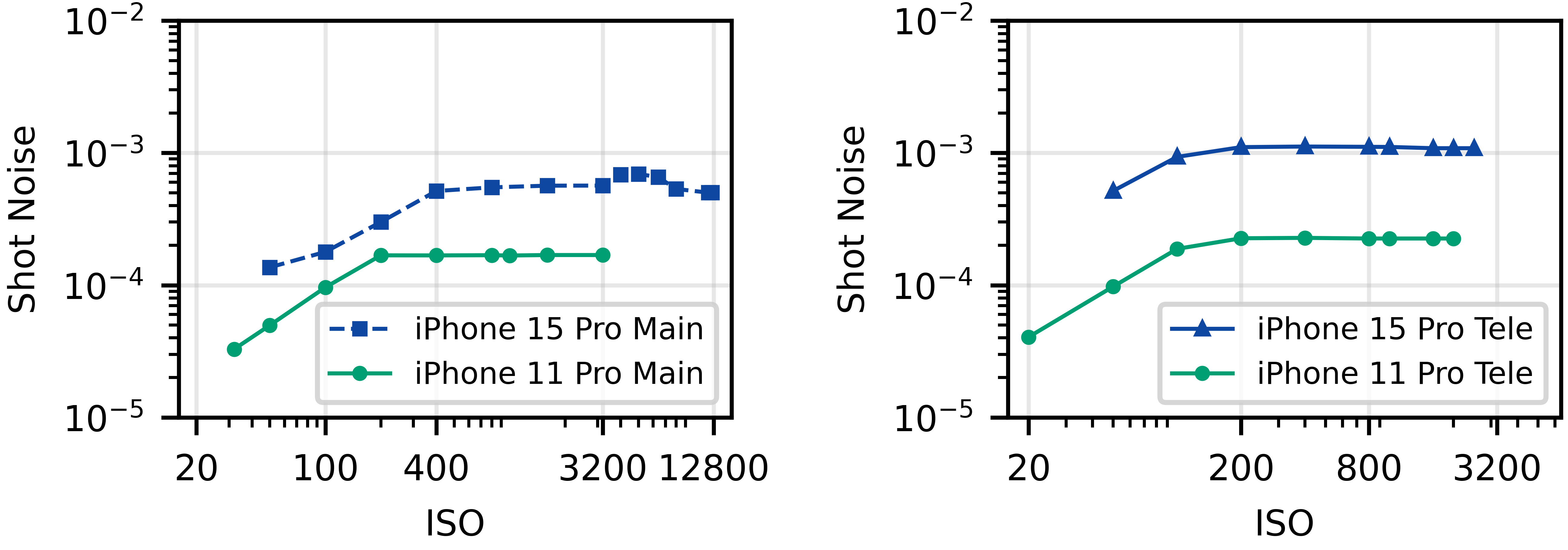}
  \caption{iPhone 11 Pro and iPhone 15 Pro DNG Profiles. Both iPhone generations exhibit non-linearity at high ISO values. The cameras have different true ISO ranges.}
  \label{fig:scaling}
\end{figure}

To verify that iPhone DNG noise profile issues are not unique to the
iPhone 11 Pro, we extracted DNG noise profiles from images captured with
the iPhone 15 Pro (released in 2023). The DNG noise profiles for the
iPhone 15 Pro also only include a single pair of values, rather than
per-channel noise parameters. \cref{fig:scaling} compares the scaling term for
both iPhone 11 Pro and iPhone 15 Pro cameras. All four DNG models
exhibit similar non-linear behavior at high ISO values, with the iPhone
15 Pro DNG model showing somewhat greater irregularities, confirming DNG noise model inaccuracies occur in multiple iPhone generations. These plateaus fail to model real-world noise, resulting in a significant gap if used to synthesize noise, and necessitate a deeper investigation, which we address in the following section.

\subsection{Non-linearity in iPhone Noise
Models}\label{non-linearity-in-iphone-noise-models}

We discovered a striking divergence between the Sony and iPhone cameras: while the Sony cameras had a fairly consistent linear relationship with ISO (in log-log space), the DNG
models for the iPhone sensors sharply plateau, which is physically
impossible. We systematically ruled out two primary explanations: calibration errors and metadata errors. When calibrating, we used the same capture process as for the Sonys, which produced valid results. To reduce the impact of the ISP, we developed an iPhone app which configured a custom AVFoundation capture pipeline disabling all optional fusion and high dynamic range features. We ruled out erroneous metadata in the DNG because our independent P-G models replicated the DNG profile's plateau. This led us to visually inspect textured RAW DNG images captured with our app. These revealed significant noise absent in the flat-field images, despite being captured with identical settings. This points to the iPhone ISP employing non-optional, content-aware smoothing to the RAW data stream. We found this necessitated the careful capture of texture-based "tuning images" to build the true noise model in order to produce iPhone images with accurately synthesized noise.

\begin{figure}[]
  \centering
  \begin{minipage}{0.35\textwidth}
    \centering
    \includegraphics[height=1.75cm]{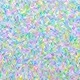}
    \caption*{Tuning Image}
  \end{minipage}
  \begin{minipage}{0.35\textwidth}
    \centering
    \includegraphics[height=1.75cm]{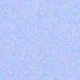}
    \caption*{Calibration Image}
  \end{minipage}
  \caption{Crops of similar flat areas in iPhone 11 Pro images (ISO 1600, 1/320s). The tuning image exhibits stochastic noise, while the calibration image is smooth, confirming iPhone ISP applies non-optional smoothing to uniform regions.}
\end{figure}

We tuned and evaluated our models using 6 scenes (3 scenes each captured with the iPhone and Sony RX100 IV, and 3
additional scenes with the Sony A7R III). Each scene
was captured at a low ISO and 6-7 higher ISO levels in powers of two. We captured noisy images in pairs for baseline noise analysis, yielding 156 full-resolution RAW tuning images
spanning most or all of the true ISO range of each camera. All images were captured using the same setup as calibration. The tuning images contained diverse textures, colors, and high-frequency content, both indoors and outdoors.

\subsection{Calibration Procedure}\label{calibration-procedure}

To increase the applicability of our findings, we chose cameras (\cref{camera-list}) spanning a range of types (smartphone, point-and-shoot, and DSLR), sensor sizes, and use cases. We also ensured they had similar focal lengths and could be tethered.

\setlength{\tabcolsep}{8pt}
\begin{table}[]
  \scriptsize
  \centering
  \caption{Cameras Used for Calibration and Synthesis}
  \begin{tabular}{l r r r}
  \toprule\noalign{}
    \textbf{Camera} & \textbf{Megapixels} & \textbf{Lens Type} & \textbf{Equiv. Focal Length} \\
   \midrule\noalign{}
    iPhone 11 Pro Main & 12 & Prime & 26 mm \\
    iPhone 11 Pro Tele & 12 & Prime & 52 mm \\
    Sony RX100 IV & 20 & Fixed Zoom & 24-70mm\\
    Sony A7R III & 42 & Zoom & 24-105mm\\
    \bottomrule\noalign{}
  \end{tabular}
  \label{camera-list}
\end{table}

For the Sony A7R III, we attached the Sony FE 24-105mm F4 G OSS lens to cover a similar focal length range as the other cameras. This lens was used for all calibration, tuning, and dataset images. The iPhones have fixed apertures; for the Sony RX100 IV and A7R III, we captured images at median apertures and focal lengths to minimize distortion and shading.

We calibrated the cameras outdoors during clear, sunny days to
ensure consistent, even lighting. The first step was to take a series of
flat-field images. In order to fill the frame, we attached a large
poster board to a sheltered, north-facing wall. We also used the
ChromLives White Balance card for some tests. All cameras were mounted
on a sturdy Manfrotto tripod and leveled in all three dimensions, and
positioned close to the poster or card while taking care to avoid
casting shadows on the subject. For the Sony cameras, we used Sony's
Imaging Edge tethering software on a Windows PC to control the camera
without touching it. For the iPhone, we developed a custom app. The iPhone app provided full control over the
ISO and exposure times, while ensuring the DNGs were as unprocessed as
Apple allows. In conjunction, we developed a macOS tethering app to
control the iPhone app over a wireless network.

\subsubsection{Flat-Field Calibration}\label{flat-field-calibration}

\begin{figure}[]
  \centering
  \begin{minipage}{0.45\textwidth}
    \centering
    \includegraphics{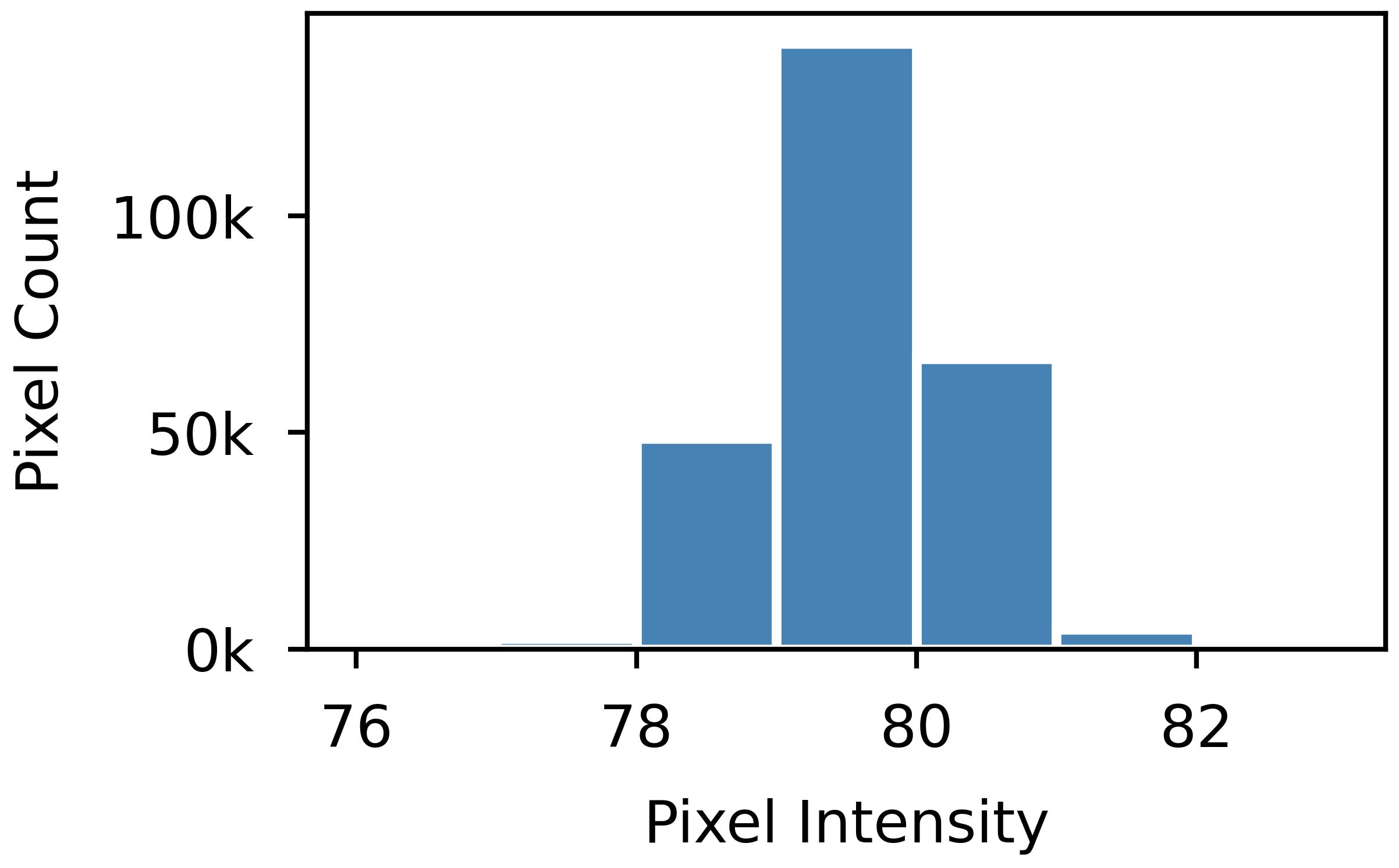}
  \end{minipage}
  \begin{minipage}{0.45\textwidth}
    \centering
    \includegraphics{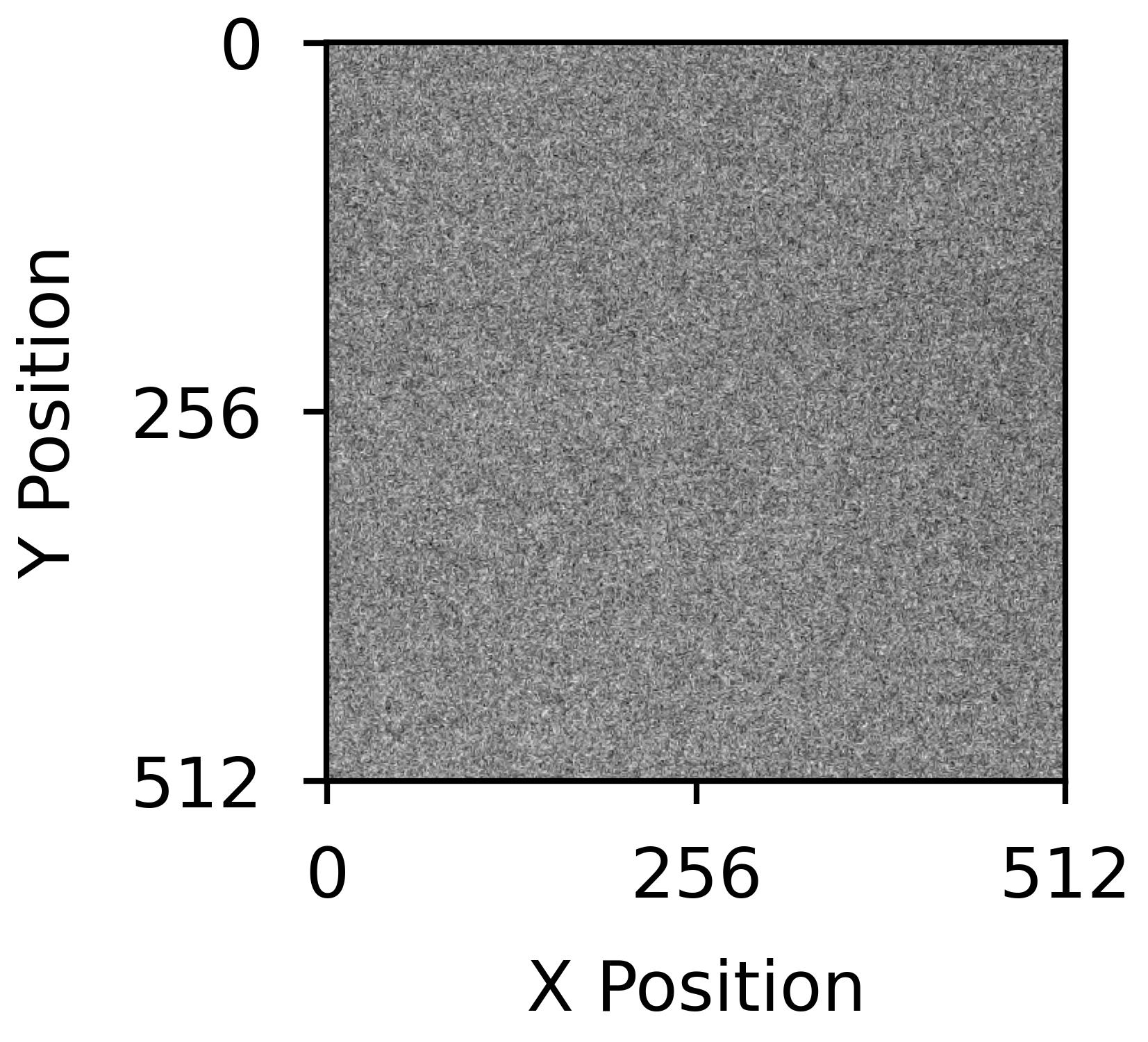}
  \end{minipage}
  \caption{Flat-field crop (512x512) for Sony A7R III showing low variance in intensity.}
  \label{flat-field-crop}
\end{figure}

We selected each camera's lowest true ISO,
focusing manually to infinity to avoid capturing texture in the poster or
card. The iPhone cameras have fixed apertures; for the
Sonys, we selected an aperture of f/8 to decrease vignetting and
a midpoint focal length to decrease distortion. We also disabled Sony's
in-camera long exposure noise reduction, and recorded uncompressed RAW
images, rather than Sony's default lossy compressed format. For all
cameras, we took pairs of RAW images at a range of ISO values and
exposure times in order to capture the full range of brightness values.
Our code examined those images to identify the area near the center of
the image that had the least variance in pixel values for calibration (\cref{flat-field-crop}).
We targeted the center region to further minimize lens distortion and
shading.

After collecting the flat-field images and identifying the crop area,
for each ISO, we grouped and sorted the images into pairs by exposure
time. The Bayer pattern data was cropped and the channel means were
computed. Images that were too dark ($\text{mean} \leq 0.02$) or too bright
($\text{mean} \geq 0.98$) were discarded to avoid clipped data. We tiled
each pair of surviving images. For each tile pair, we computed the tile
means and the variance by subtracting one tile from its
partner. To estimate the Poisson-Gaussian noise parameters, we used a
two-stage procedure: first, a HuberRegressor identified and removed
outliers from the mean-variance relationship; second, a standard
least-squares linear regression on the resulting cleaned data yielded
the final Poisson-Gaussian noise model parameters (photon shot noise
coefficient $a$ and read noise variance $b$).

\begin{figure}[]
  \centering
    \includegraphics[height=3cm]{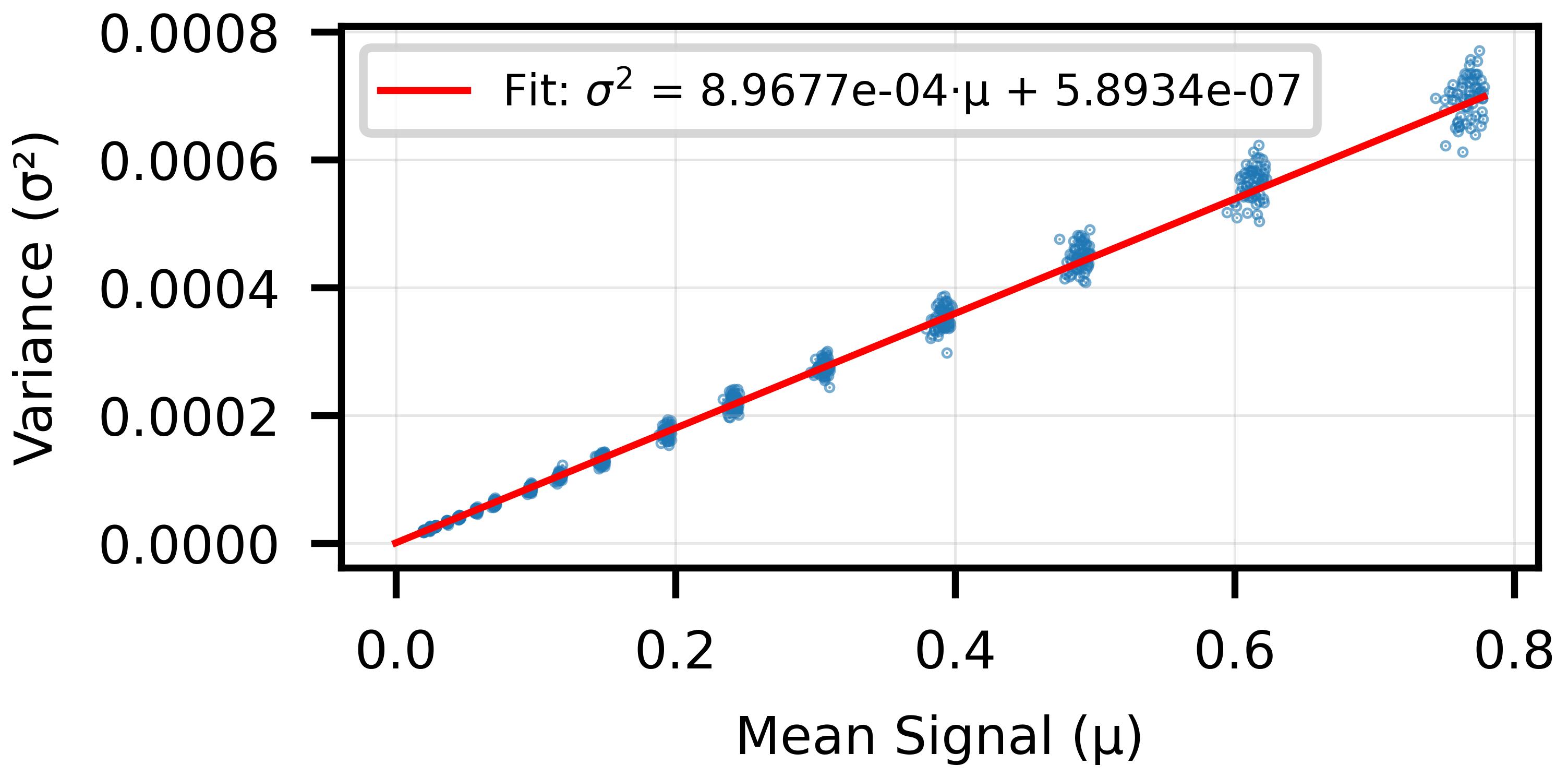}
  \caption{Photon Transfer Curve (Red) for iPhone 11 Pro main camera at ISO 3200.}
  \label{fig:ptc}
\end{figure}

We built Photon Transfer Curves (PTC) for all cameras and ISO values
using the red channel. \cref{fig:ptc} shows a representative plot at ISO 3200.
The red line indicates the fitted P-G model (\(\sigma^2{} = a\mu + b\)),
with R\textsuperscript{2}=0.9995. The parameters $a$ and $b$ from
this fit were then used in the P-G synthesis model. The increase in
spread of measured points as mean signal rises is consistent with photon
shot noise, while the presence of outliers at the higher signal levels
demonstrates the need for robust regression techniques.

\subsubsection{Dark Current Capture}\label{dark-current-capture}

Read noise can be difficult to fully characterize in modern
sensors \cite{Zhang}. While flat-field calibration can characterize read
noise, a more robust technique uses dark frames. Dark frames are
captured with a lens cap when possible, and in a dark room or
environment. We placed cameras face down inside black fabric pouches and
placed those inside sealed boxes, with a small hole cut for the
tethering cable. Then per the technique in Zhang \etal~\cite{Zhang}, we captured 10 dark
frames at each ISO at an exposure time of 1 second. We took care to
disable any in-camera noise reduction to get accurate dark
frame images. Those dark frame images were then used in the synthesis process.

\subsection{Tuning the Noise Model}\label{tuning-the-noise-model}

\begin{figure}[t]
    \centering
    \begin{subfigure}[b]{0.24\textwidth}
        \centering
        \includegraphics[height=2cm]{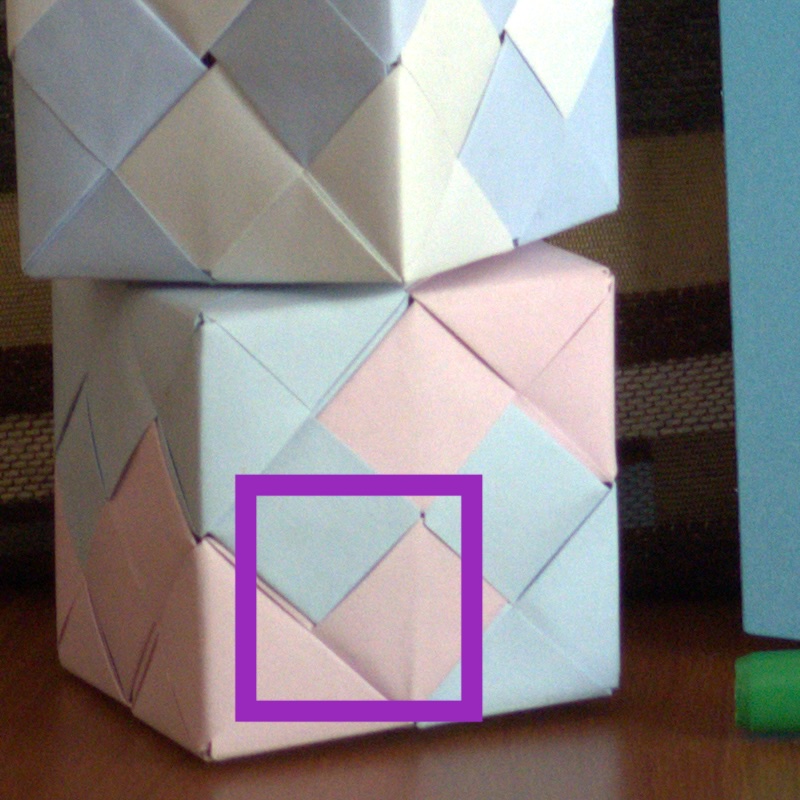}
        \caption{Clean Image Crop}
    \end{subfigure}
    \hfill
    \begin{subfigure}[b]{0.24\textwidth}
        \centering
        \includegraphics[height=2cm]{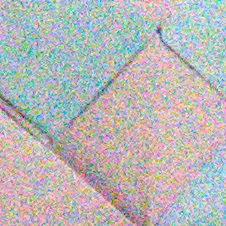}
        \caption{Real Noisy Image}
    \end{subfigure}
    \hfill
    \begin{subfigure}[b]{0.24\textwidth}
        \centering
        \includegraphics[height=2cm]{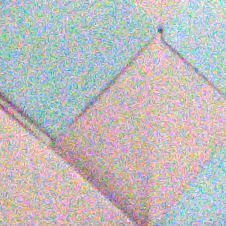}
        \caption{Ours (Tuned)}
    \end{subfigure}
    \hfill
    \begin{subfigure}[b]{0.24\textwidth}
        \centering
        \includegraphics[height=2cm]{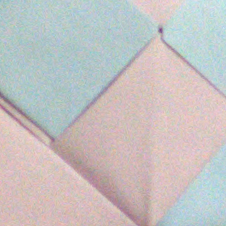}
        \caption{DNG Noise Model}
    \end{subfigure}
  \caption{Comparison of noise at ISO 1600 for the iPhone 11 Pro telephoto camera.}
  \label{fig:crops}
\end{figure}

We observed non-linearity in the initial noise models. The iPhones showed extensive non-linearity, while the Sony RX100 IV was only affected at the highest ISOs. Because ISO scales geometrically, we fit a power-law relationship to the P-G shot noise parameter $a$ using textured tuning images. For the Sony RX100 IV, this refined the model to match the sensor's linear response (in log-log space). For the iPhone, it compensated for the sensor's content-aware smoothing identified during calibration. We constrained per-channel fits to two middle-range ISO values, as these points span a sufficient range of ISOs while avoiding non-linearities at extreme ISOs ($R^2 > 0.999$). The relationship follows: 
\begin{equation}
a = k \cdot \text{ISO}^m,
\end{equation}
where $k$ and $m$ are channel-specific
constants determined by the two middle-range ISO values. We performed this tuning
for the iPhone sensors and for the Sony RX100 IV; the professional-quality Sony A7R III required none. Our tuning improvements resulted in superior LPIPS
measurements vs.~the images that were calibrated but not tuned. The measurable benefits of tuning are evaluated in the Results section. \cref{fig:crops} shows full-size crops of real noise, DNG model noise, and our tuned and calibrated noise for the iPhone 11 Pro telephoto lens at ISO 1600.

\subsection{Noise Synthesis}\label{noise-synthesis}

We employed noise synthesis at two different stages: first to generate noisy images for model evaluation and tuning, and second to create SNIC dataset images by adding synthesized noise to a separate set of clean low-ISO dataset images. For each target noisy ISO, we loaded the calibrated per-channel (and per-ISO) noise model and the database of dark frame images. For the iPhone sensors, using dark-current databases rather than the Gaussian $b$ term reduced the real-to-synthetic gap (confirming Zhang \etal~\cite{Zhang}). Conversely, for the high-performance Sony sensors, our calibrated P-G model was highly accurate with negligible read noise. Adding dark frame sampling provided no measurable improvement in synthesis accuracy and was therefore omitted for these models for computational efficiency.

Given a clean low-ISO RAW image Bayer pattern channel
\(I_{\text{clean}}\) normalized to \([0,1]\), we synthesized and injected
Poisson-Gaussian noise as follows:
\begin{equation}
\lambda_{\text{noisy}} \sim \mathcal{P}\left(\frac{I_{\text{clean}}}{a}\right),
\end{equation}
\begin{equation}
I_{\text{noisy}} = a \cdot \lambda_{\text{noisy}} + n_{\text{read}}, \quad \text{where } n_{\text{read}} \sim \mathcal{N}(0, b),
\end{equation}
where $a$ is the photon-to-digital-number conversion factor, and $b$
is the read noise variance. This process was applied to all four color
channels (R, G1, G2, B) of the RAW Bayer pattern. This approach was used
for cameras that did not benefit from dark current synthesis (the Sony
cameras in our case).

When incorporating dark current, we ignored the $b$ term from the P-G noise model. Following Zhang \etal~\cite{Zhang}, we randomly selected one of ten dark frames and injected dark current with spatial alignment maintained to increase the accuracy of the synthesis. Using the final noisy Bayer data, we produced RAW DNG files and 16-bit TIFFs and grouped them with the low-ISO clean images, arranged by camera and scene.

\section{Results}\label{results}

\subsection{Evaluating Synthesis
Quality}\label{evaluating-synthesis-quality}
\begin{figure}[]
  \centering
  \includegraphics[width=1.2in]{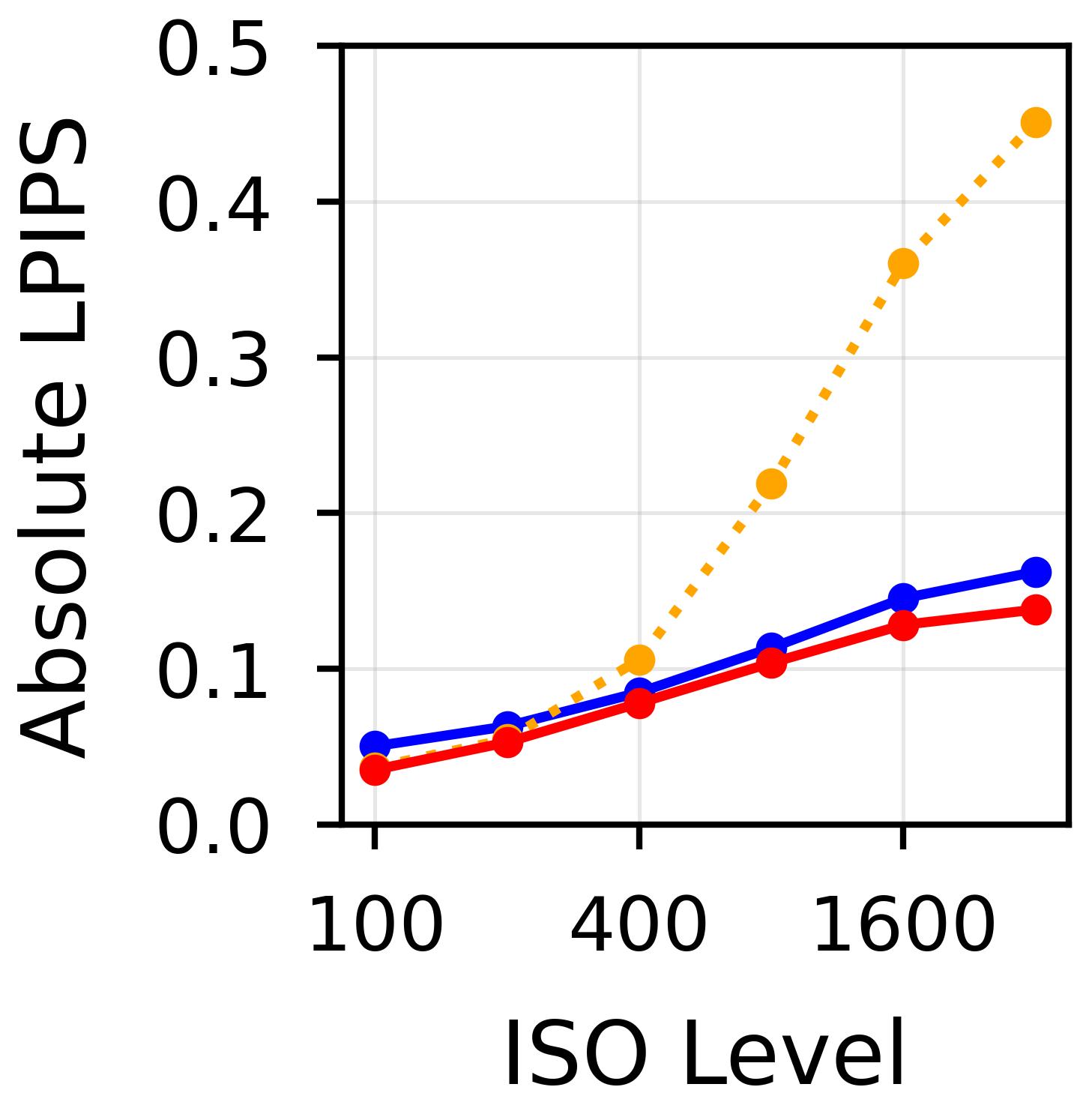}%
  \includegraphics[width=1.2in]{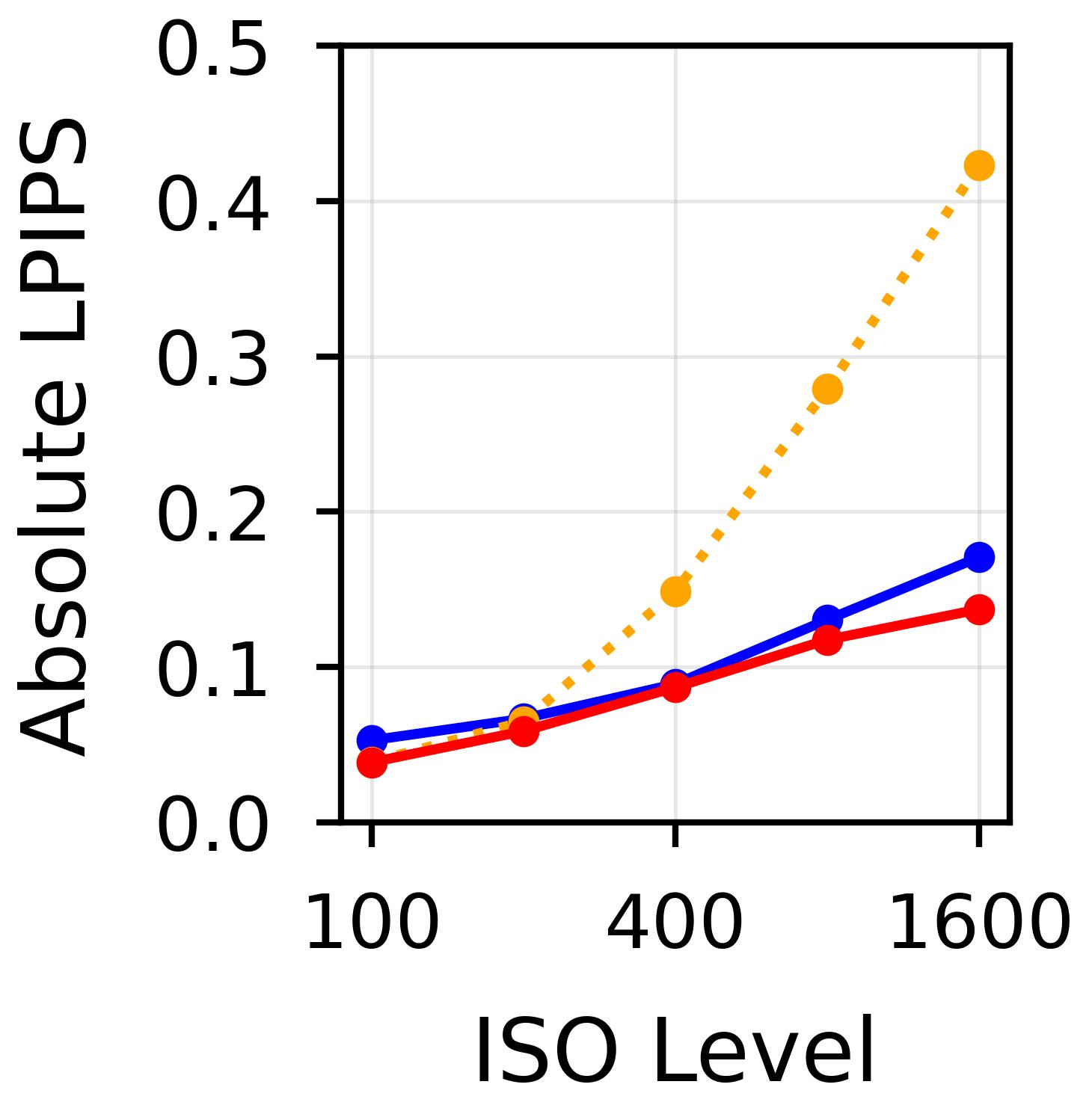}%
   \includegraphics[width=1.25in]{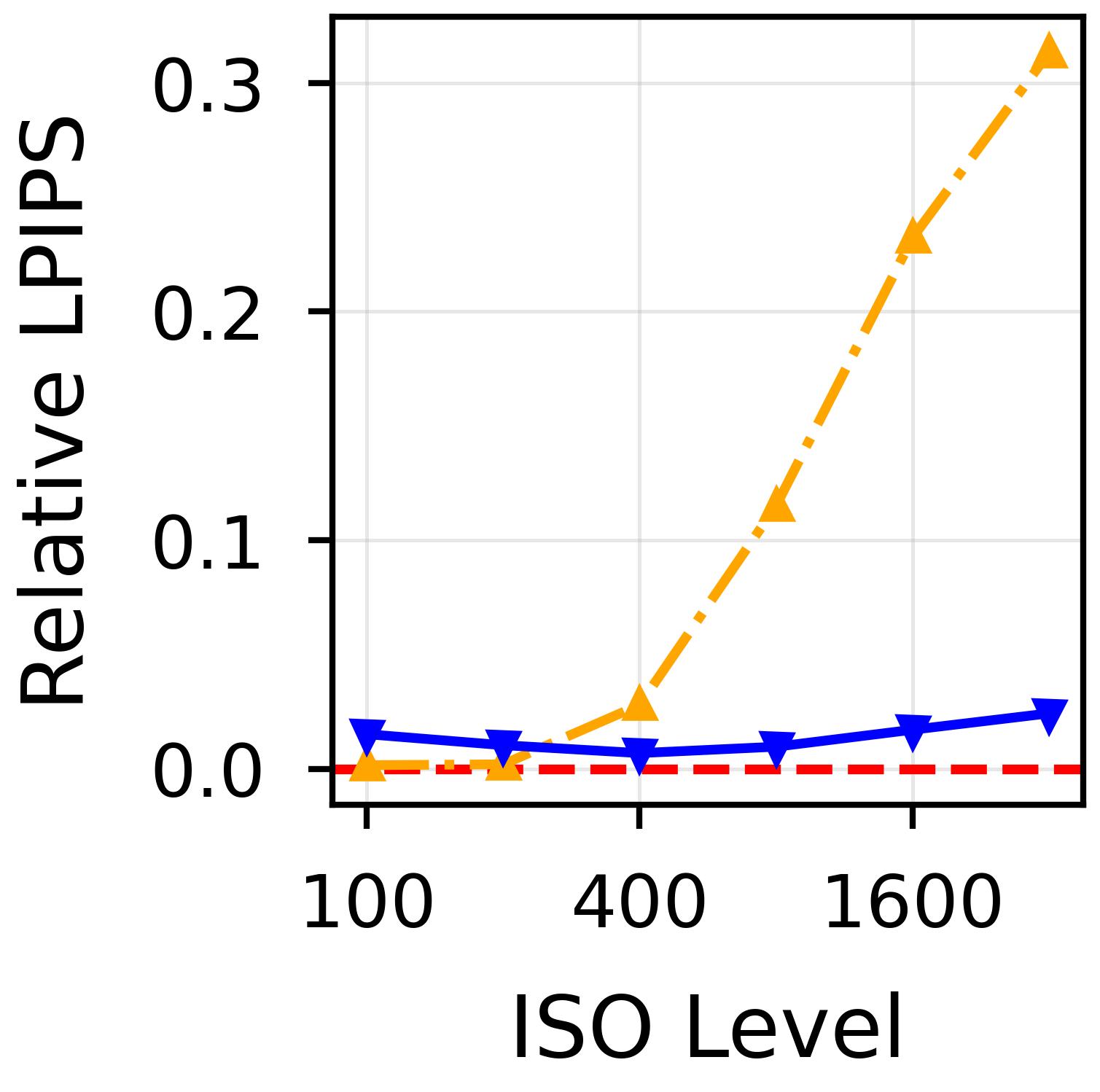}%
 \includegraphics[width=1.25in]{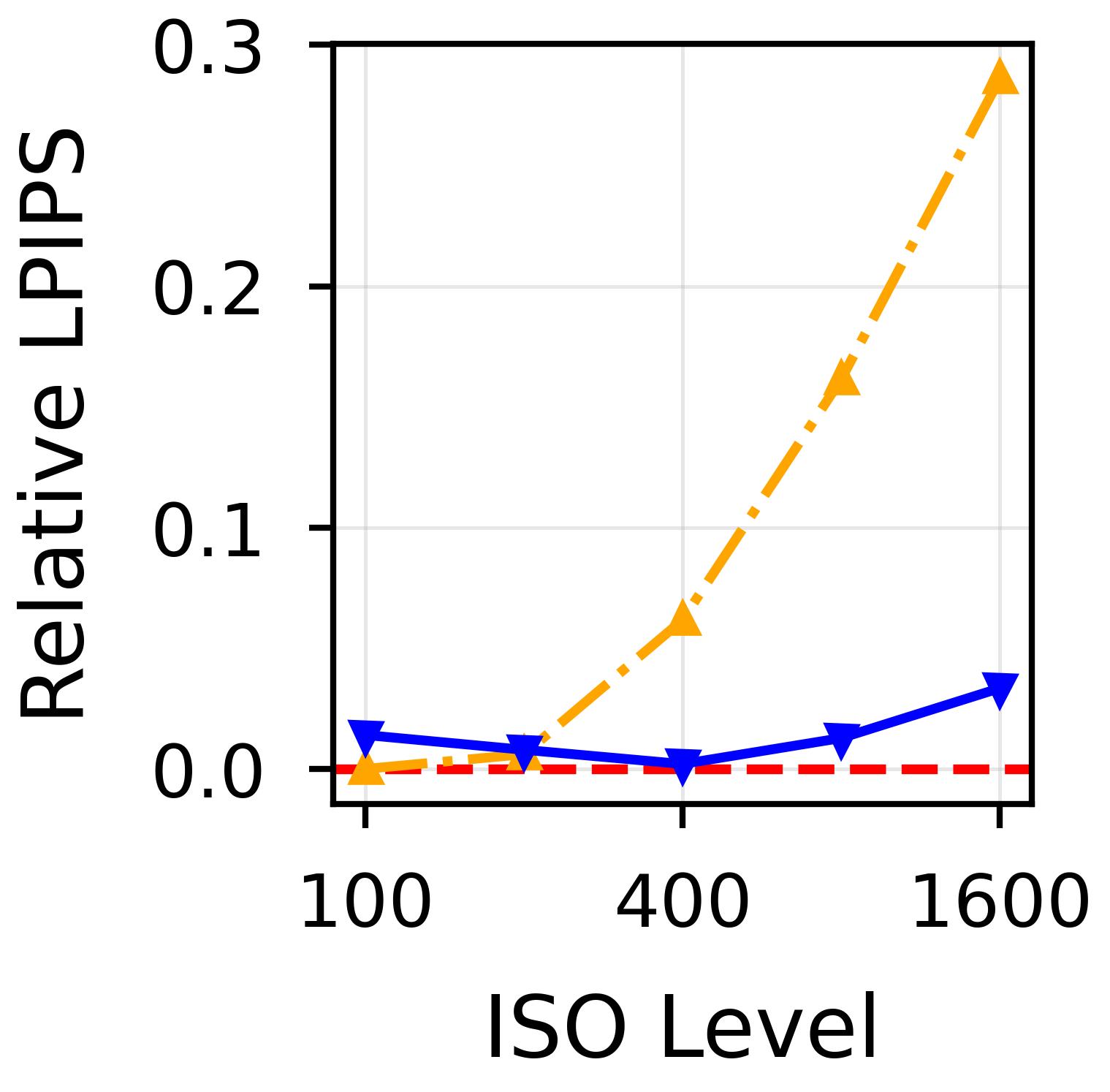}\\[2pt]
\begin{minipage}{1.2in}
  \centering  \enspace\enspace  iPhone Main
\end{minipage}%
\begin{minipage}{1.2in}
  \centering  \enspace\enspace iPhone Tele
\end{minipage}%
\begin{minipage}{1.25in}
  \centering  \enspace\enspace iPhone Main
\end{minipage}%
\begin{minipage}{1.25in}
  \centering  \enspace\enspace iPhone Tele
\end{minipage}\\[4pt]
\begin{minipage}{2.8in}
  \centering
  \includegraphics[width=2.7in]{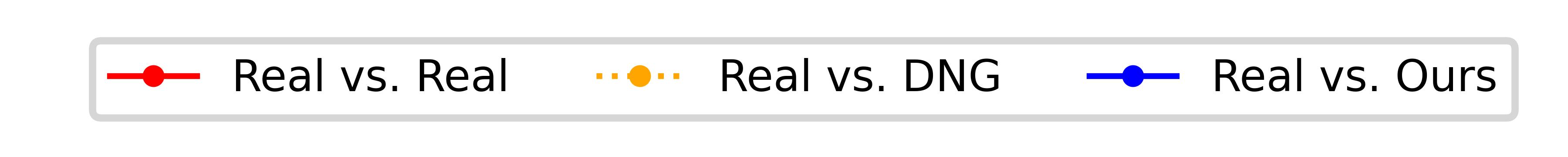}
\end{minipage}%
\begin{minipage}{2.3in}
  \centering
  \includegraphics[width=2.2in]{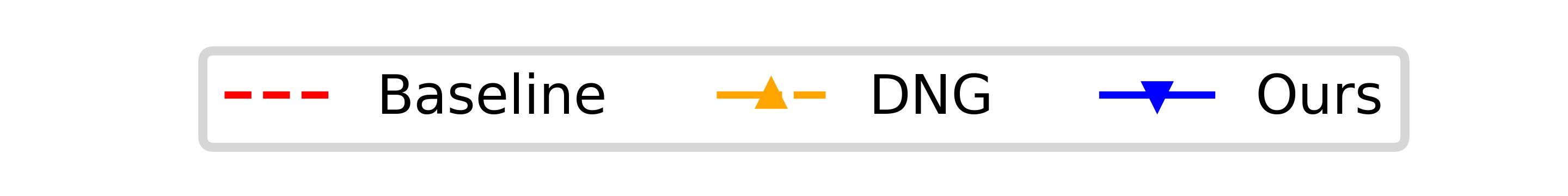}
\end{minipage}
  \caption{Absolute LPIPS (left pair) and Relative LPIPS (right pair). Relative LPIPS is computed against a real-vs-real noisy baseline placed at 0.0 in the relative plots.}
  \label{fig:iPhone11_lpips}
\end{figure}

We compared synthesized noisy images to real ones captured in pairs at each ISO during tuning. We considered three metrics: LPIPS~\cite{LPIPS}, SSIM~\cite{SSIM}, and PSNR. PSNR measures pixel-level differences, making it unsuitable for comparing different realizations of noise: applied to real noisy image pairs, PSNR ranged from -69 dB to below -80 dB. For SSIM at high ISO values, noise dominates local window statistics, yielding values near 0.0 (no similarity). In contrast, LPIPS, a learned perceptual similarity metric, consistently identified real noisy pairs as similar, and is therefore our metric for comparing noisy images. We plotted both absolute values and values relative to a real-vs-real baseline for each sensor. Lower absolute values and relative values closer to zero indicate more realistic synthesis. \cref{fig:iPhone11_lpips} shows absolute and relative LPIPS results for the iPhone cameras. The calibrated and tuned noise are clearly superior to DNG model noise and close to real noisy images, even at higher ISO values. LPIPS plots for the Sony cameras are omitted as the improvement is modest and consistent with the quantitative results in Table 2. In the following section, we used PSNR to compute the real-to-synthetic gap when denoising images with NAFNet.

\subsection{NAFNet Denoising Results}\label{nafnet-denoising-results}

To evaluate the realism of our noise models, we used the NAFNet \cite{NAFNet} denoising model, which achieves state-of-the-art results. We used the NAFNet-SIDD-width64 model, pretrained on the SIDD dataset \cite{SIDD} and not fine-tuned, ensuring NAFNet has not seen any of our images or sensors and the results reflect the accuracy of our synthesis rather than overfitting or dataset leakage. We constructed testing datasets for each
camera with three noise approaches: real noise, DNG model, and our tuned noise model, each with a low-ISO ground truth image. The SIDD validation set uses 256×256 pixel patches, so we generated 256×256 patches from our full-resolution tuning images. Our resulting validation set comprises 79164 patch pairs --- over 60$\times$ more than SIDD's 1280 validation patches.

The purpose of this experiment was to evaluate how well NAFNet generalizes to different noise
types (real, DNG model, and ours) using identical scenes, cameras, and
ISO values, rather than comparing absolute performance with SIDD. Using
PSNR and SSIM metrics, NAFNet denoised our noisy images
significantly more accurately than noisy images created using manufacturer DNG models.
For the iPhone main camera, the PSNR gap between our synthesized noise
and real noise was 1.89 dB versus 5.33 dB for DNG model synthesis --- a
64.5\% improvement. The SSIM gap was 0.083 versus 0.141, a 41.1\%
improvement. These results demonstrate that our calibrated iPhone models
produce substantially more realistic noise than manufacturer DNG models. For the Sony A7R III, there was strong agreement between our model and Adobe's noise profile. This agreement serves as independent validation of our pipeline and methods: reproducing Adobe's independently-derived model would be unlikely if our methods introduced calibration errors. We conclude that our iPhone noise models correct a systemic ISP effect, rather than compensating for errors in our procedure.

\setlength{\tabcolsep}{3pt}
\begin{table}[t]
\centering
\scriptsize
\caption{Quantitative results from NAFNet (PSNR/SSIM) comparing synthesis methods across cameras. Real-world noise serves as the baseline. DNG Model represents Poisson-Gaussian models from DNG noise profiles (manufacturer or Adobe-provided). Smaller gaps indicate more realistic synthesis. Best synthesis methods per metric are emboldened and highlighted in \textcolor{blue}{blue}. "Pairs" refers to 256x256 validation tiles.} 
\begin{tabular}{llrrr|rr|r}
\toprule
Camera & Metric & Real & DNG & Ours & Gap (DNG) & Gap (Ours) & Improvement \\
\midrule
\multirow{2}{*}{\shortstack[l]{iPhone 11 Pro Main \\ {\scriptsize (10,368 pairs)}}}
& PSNR & 25.91 & 31.24 & \textcolor{blue}{\textbf{27.80}} & 5.33 & \textcolor{blue}{\textbf{1.89}} & 64.5\% \\
& SSIM & 0.799 & 0.940 & \textcolor{blue}{\textbf{0.882}} & 0.141 & \textcolor{blue}{\textbf{0.083}} & 41.1\% \\
\midrule
\multirow{2}{*}{\shortstack[l]{iPhone 11 Pro Telephoto \\ {\scriptsize (8,640 pairs)}}}
& PSNR & 27.48 & 31.21 & \textcolor{blue}{\textbf{29.21}} & 3.73 & \textcolor{blue}{\textbf{1.73}} & 53.6\% \\
& SSIM & 0.850 & 0.936 & \textcolor{blue}{\textbf{0.899}} & 0.086 & \textcolor{blue}{\textbf{0.049}} & 43.0\% \\
\midrule
\multirow{2}{*}{\shortstack[l]{Sony RX100 IV \\ {\scriptsize (17,820 pairs)}}}
& PSNR & 32.57 & 32.93 & \textcolor{blue}{\textbf{32.86}} & 0.36 & \textcolor{blue}{\textbf{0.29}} & 19.4\% \\
& SSIM & 0.911 & 0.913 & \textcolor{blue}{\textbf{0.911}} & 0.002 & \textcolor{blue}{\textbf{0.000}} & 100.0\% \\
\midrule
\multirow{2}{*}{\shortstack[l]{Sony A7R III \\ {\scriptsize (42,336 pairs)}}}
& PSNR & 36.59 & 37.95 & \textcolor{blue}{\textbf{37.90}} & 1.36 & \textcolor{blue}{\textbf{1.31}} & 3.7\% \\
& SSIM & 0.961 & 0.966 & \textcolor{blue}{\textbf{0.966}} & \textcolor{blue}{\textbf{0.005}} & \textcolor{blue}{\textbf{0.005}} & 0.0\% \\
\bottomrule
\end{tabular}
\label{tab:nafnet}
\end{table}

\subsection{Ablation Study}\label{ablation-study}
\begin{figure}[]
  \centering
    \includegraphics{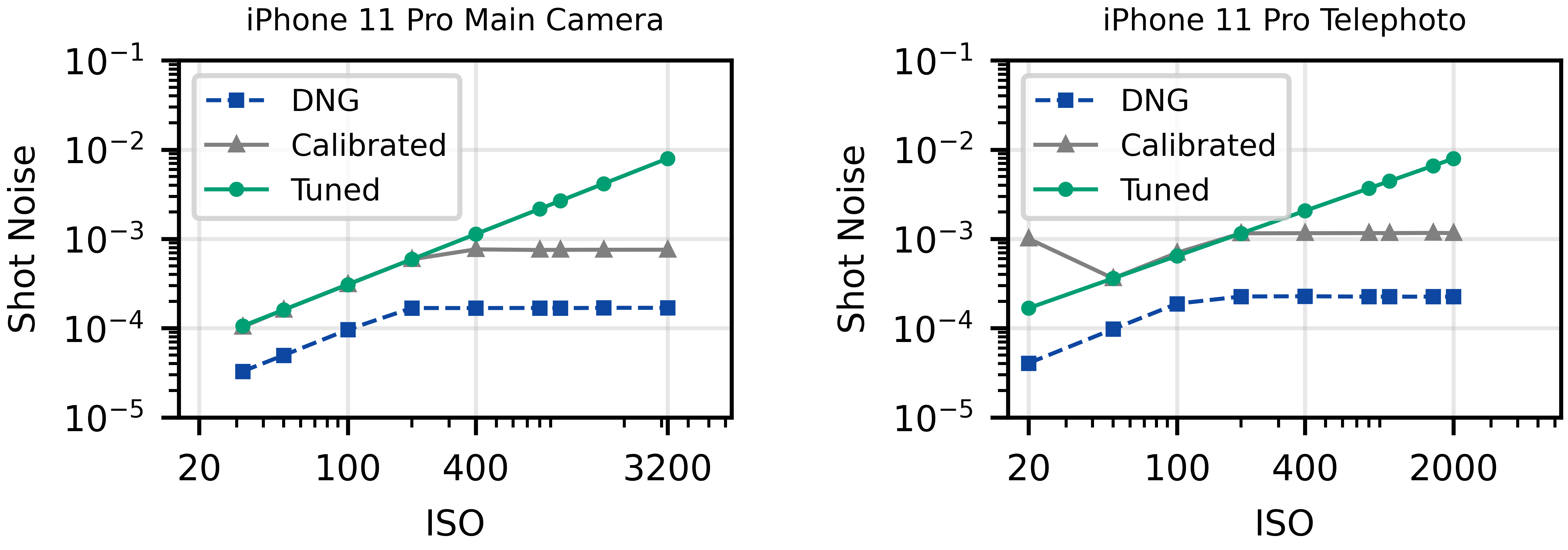}
  \caption{DNG, Calibrated, and Tuned Noise Models for the iPhone 11 Pro.}
  \label{fig:dng}
\end{figure}

To isolate the contributions of calibration and tuning, we compared three models (DNG profiles, calibrated only, and calibrated-and-tuned) holding the scenes, cameras, and camera parameters constant. In \cref{fig:dng}, the DNG model exhibits clear non-linearity. Calibration establishes a more accurate baseline, but retains the ISP-induced plateau. Tuning fully resolves the plateau and corrects measurement anomalies at extreme ISO values. As shown in \cref{evaluating-synthesis-quality} and \cref{nafnet-denoising-results}, tuned models produce noise that is measurably more realistic, both perceptually (LPIPS) and when denoised with the Nonlinear Activation Free Network (NAFNet \cite{NAFNet}). The greatest improvements were shown with iPhone cameras whose DNG profiles had the most severe inaccuracies.

\section{Dataset}\label{dataset}

In addition to building accurate heteroscedastic noise models, we have created the only known public dataset of high-fidelity RAW images with synthesized noise. The images are perfectly aligned by construction, and the sensors span a range from mobile to professional, ensuring the dataset captures a wide spectrum of noise behaviors.

\setlength{\tabcolsep}{8pt}
\begin{table}[h!]
  \scriptsize
  \centering
  \caption{Synthesized Noisy Images using Calibration dataset (SNIC)}
  \begin{tabular}{l r r r r}
  \toprule\noalign{}
    \textbf{Camera} & \textbf{Clean} & \textbf{Total} & \textbf{ISOs} & \textbf{Size (GB)} \\
   \midrule\noalign{}
    iPhone 11 Pro Main & 132 & 1848 & 32-3200 & 79.28 \\
    iPhone 11 Pro Tele & 143 & 2002 & 20-1600 &  85.68 \\
    Sony RX100 IV & 79 & 1264 & 125-12800 &  92.47\\
    Sony A7R III & 93 & 1488 & 100-12800 &  229.63 \\
   \midrule\noalign{}
    \textbf{Totals} & \textbf{447} & \textbf{6602} & & \textbf{487.06}\\
    \bottomrule\noalign{}
  \end{tabular}
\end{table}
For each sensor, we captured the same 30 scenes (15 indoor and 15
outdoor), which are distinct from the ones used for tuning, and cover a range of locations, lighting, colors, and textures. All images were captured in RAW format at the sensor's lowest real ISO. 
We injected noise for a range of ISO values using our tuned noise models into the low-ISO RAW original, using the dark
frame database when appropriate. For each camera and scene, we included the low-ISO RAW image and a rendered 16-bit sRGB TIFF, plus pairs of noisy RAW DNGs and noisy 16-bit sRGB TIFFs at each synthesized ISO value. RAW images were decoded using LibRaw (rawpy), with auto-brightness and auto-scaling disabled during noise modeling, and enabled for TIFF generation to produce natural-looking output. A comprehensive datasheet following the Gebru framework \cite{Gebru} is included in the supplement to provide documentation transparency.

\subsection{Limitations}\label{limitations}
Our noise models are specific to the four cameras used, though the code, calibration, and tuning methodologies are generic and adaptable to other hardware. However, creating additional high-fidelity noise models is a rigorous and labor-intensive process. It requires careful lighting, precise configurations, and stable environmental conditions. Additionally, iPhone ISPs are "black boxes" which may vary in other shooting modes or due to software updates, potentially affecting the ISP's content-aware smoothing and our noise modeling findings.

\section{Conclusion}\label{conclusion}

In this paper, we describe the advantages of synthesizing noisy images using calibrated and tuned noise models, in conjunction with dark current frames. Our approach produces images that are quantitatively superior to both manufacturer and Adobe noise models. We examined our images visually, used the LPIPS metric to compare them with real noise, and measured PSNR after denoising our images with NAFNet. Our images outperformed published DNG Poisson-Gaussian models, reducing the PSNR gap by 54–64\% for iPhone sensors. Our models were created per ISO and per channel, using carefully captured calibration images, and sound mathematical modeling. The result is a publicly available dataset that, to our knowledge, provides the only collection of calibrated and tuned noise in both RAW and TIFF formats from a range of sensors and with a diverse collection of indoor and outdoor scenes. Researchers can use it for training, or use our code to create their own.


\section*{Acknowledgments}
Some of Kaiser's other research is supported in part by NSF CNS-2247370 and NSF CCF-2313055. Any opinions, findings, conclusions or recommendations expressed herein are those of the authors and do not necessarily reflect those of the US Government or NSF.

%
%
\bibliographystyle{splncs04}
\bibliography{references}
\end{document}